# Josephson Current and Multiple Andreev Reflections in Graphene SNS Junctions

Xu Du, Ivan Skachko and Eva Y. Andrei
Department of Physics and Astronomy, Rutgers University, Piscataway, New Jersey  08855

Abstract

The Josephson Effect and Superconducting Proximity Effect were observed in Superconductor - Graphene-Superconductor (SGS) Josephson junctions with coherence lengths comparable to the distance between the superconducting leads. By comparing the measured temperature and doping dependence of the supercurrent and the proximity induced sub-gap features (multiple Andreev reflections) to theoretical predictions we find that, contrary to expectations, the ballistic transport model fails to describe the SGS junctions. In contrast, the diffusive junction model yields close quantitative agreement with the results. This conclusion is consistent with transport measurements in the normal state, which yield mean free paths in the graphene link that are much shorter than the junction length.  We show that all devices fabricated on $SiO_2$ substrates so far (our own as well as those reported by other groups) fall in the diffusive junction category.

The discovery of methods to extract single atomic layers from graphite[1, 2] (graphene) has triggered a torrential effort to explore the new physical properties emerging from their relativistic (Dirac) quasiparticle spectrum[2-4]. A particularly interesting set of questions and expectations has arisen with the recent fabrication of graphene-superconductor (GS) hybrid structures[5, 6], which has made it feasible to study the interplay between superconductivity and relativistic quantum dynamics. Because of the chemical inertness of graphene, achieving transparent interfaces is relatively easy and reproducible compared to other gate controllable junctions where the weak link is a semiconductor or a 2D electron gas[7, 8]. With almost ideal interfaces, and the ability to carry bipolar supercurrents that are gate tunable from electron to the hole branch[5, 9], the SGS junctions are promising candidates for nano-electronics applications as well as for studying the physics and "phase diagram" of Josephson junctions[10]. It is therefore important to understand the basic properties of experimentally realizable SGS junctions. These properties are expected to be controlled by the transport of relativistic electrons across the GS interface which is qualitatively different from the transport of normal electrons. Whereas a normal electron impinging on a GS interface is "retro-reflected" as a hole (Andreev reflection) retracing the same trajectory[11-13], the process is specular for relativistic electrons[14] (if the Fermi energy is within the superconducting gap). These "specular Andreev reflections" (SAR) are expected to leave clearly manifest marks in ballistic SGS junctions, where the electron mean free path exceeds the junction length, detectable through a strong and unusual gate dependence of the Multiple Andreev Reflections (MAR) [11-13,15]. Furthermore, in ballistic SGS junctions the Josephson critical current, $I_c$, and the product $I_cR_n$ ($R_n$ is the normal state resistance) are expected to exhibit a characteristic gate dependence, which is qualitatively different from that of conventional SNS junctions[9].

Many proposed physical phenomena and devices based on SGS junctions implicitly assume ballistic transport. This is because, due to unique properties such as chirality and Zitterbewegung[16, 17], the relativistic carriers in graphene are expected to be rather insensitive to scattering and to have long mean free paths. Surprisingly, thus far, there is no solid experimental evidence in support of ballistic transport or of relativistic charge carriers in SGS. Here we show that SGS junctions fabricated on Si/SiO$_2$ substrates with current techniques are in fact diffusive with mean free path much shorter than the junction length. In these junctions, we find that $I_c$ is more than an order of magnitude below the value predicted by the ballistic model. Furthermore, from the gate dependence of $I_cR_n$ and that of the MAR conductivity maxima, we demonstrate quantitative agreement with models of diffusive junctions.

SGS junctions were fabricated with mechanically exfoliated single layer graphene[1] deposited onto Si(p++)/SiO$_2$(300nm) substrates that were pre-deposited with alignment marks. Following the identification of the graphene layers with a combination of optical imaging and AFM, the leads, Al(30nm)/Ti(2nm), were fabricated using standard e-beam lithography and lift-off techniques. To avoid contamination of the interfaces, only optical imaging was used for the e-beam positioning and pattern design. Lead distances were in the range $L \sim 200 - 400$ nm and the aspect ratios W/L ~ 10 -30, where $W$ is the junction width. An optical image of a typical device is shown in the inset of Fig. 1a. Measurements were carried out in a dilution refrigerator with base temperature of 100 mK. Noise filtering, an essential requirement for accessing the intrinsic properties of the SGS junctions, was accomplished with two sets of filters: RC filters (2-stage with cut-off frequency of 1kHz) at low temperature (4 K) and a bank of pi-filters (Spectrum Control, ~70dB at 200MHz~2GHz) at room temperature. The transport measurements, carried out with a standard 4 lead technique, employed a commercial current source (Keithley 2400) and a lock-in amplifier (SR830) for the dV/dI measurements and a K2001 voltmeter with PAR 113 amplifier for the DC measurements. A back-gate voltage, $V_g$, applied to the Si substrate was used to control the doping level of the graphene layer, $n \sim 7.4 \times 10^{10} |V_g| \, cm^{-2}$, where $n$ is the carrier density.

Upon cooling sufficiently far below the critical temperature of the leads, $T_c \sim 1$ K, the IVC show sharp switching between Josephson and normal currents as illustrated in Fig. 1a. For convenience, unless specified otherwise, we will show detailed data only from sample S032007, with length $L = 350$ nm and width $W = 9$ μm. Most data shown here pertain to this sample, but the other samples (5 samples were measured) exhibit similar behavior. The sharp features in the IVC become smeared on approaching $T_c$. They are hysteretic, with the transition from Josephson to normal state always occurring at higher current, as expected for under-damped Josephson junctions in the RCSJ model[18]. The switching is very sensitive to magnetic field, as illustrated by the Fraunhoffer pattern dependence of $I_c$ shown in the lower inset of Fig 1a. To measure the value of $I_c$ in zero-field, we apply a compensating field tuned to maximize its value. Another type of switching induced by sweeping the back-gate voltage $V_g$ (doping level), is illustrated in Fig. 1b. As before, sharp switching is seen between the Josephson and the normal current states, this time as a function of $V_g$. Here too we observe hysteresis as a function of doping level. In the RCSJ model, both cases correspond to run-away of the "phase particle" moving in a tilted washboard potential $U(\varphi) = -E_J \left( cos(\varphi) + \frac{I}{I_c} \varphi \right)$ with average slope ~$I/I_c$, where $\varphi$ is the phase difference between the two superconducting banks, and $E_J = \frac{\Phi_0 I_c}{2\pi}$ is the Josephson energy. The slope is controlled by $I$ or by $I_c$ for the current or gate swept measurements respectively. Fig. 1c illustrates the variation of $R_n$, the resistance in the normal state, as $V_g$ is swept through the

Dirac point and the carriers change continuously from holes (negative $V_g$) to electrons (positive $V_g$). The low temperature normal state was accessed by quenching the superconductivity in the leads with a small magnetic field. The inset shows the gate dependence of the mean free path, $l$, estimated from the measured normal state conductivity, $\sigma$, using $l = \dfrac{\sigma h}{2e^2 k_F}$ and $k_F = \sqrt{\dfrac{\varepsilon \varepsilon_0 V_g \pi}{ed}}$. Here $d=300$ nm is the thickness of the SiO$_2$ spacer and $\varepsilon \sim 4$ its dielectric constant. It was recently shown that the carrier density near the DP breaks up into electron and hole puddles[19, 20] possibly as a result of the random potential and un-intentional doping created by trapped charges. As a result, the carrier density at low gate voltages remains finite, making it difficult to estimate $k_F$. At large $V_g$, where the effect of the random potential is insignificant, the carrier density can be reliably estimated and the mean free path is found to depend only weakly on doping level. Its value for all SGS samples fabricated thus far (our own as well as those reported by other groups)[5, 6] is surprisingly low $l \sim 25\,nm \ll L$, indicating that the SGS junctions are diffusive. The weak gate dependence of the mean free path suggests that scattering is dominated by defects or impurities[21, 22]. In this case, the mean free path near the DP cannot be significantly different, and the apparent divergent behavior is an artifact arising from the division by $V_g$ while the actual effective carrier density saturates upon the formation of the electron – hole puddles.

The doping dependence of $I_c$ obtained by monitoring the *IVC* while continuously varying $V_g$ is shown in Figure 1d. We find $I_c \sim 320$ nA at the DP and 1.5 µA at $V_g = 40$ V. These values, regardless of doping level, are more than an order of magnitude lower than theoretical estimates in ballistic SGS junctions[9]: $I_c(V_g = 0) = 1.33 \dfrac{e\Delta_0}{\hbar} \dfrac{\pi W}{L} \sim 3.3\,\mu A$ and

$$I_c(V_g = 40V) = 1.22 \frac{e\Delta_0}{\hbar} \frac{\mu W}{\pi \hbar v} \sim 30\mu A$$ ( for $T = 0$ and $W/L \sim 26$). This, together with the normal state transport result that $l \ll L$, suggests that the junctions should be treated as diffusive. Thus, we use the diffusive limit to estimate the superconducting coherence length:

$$\xi \sim \sqrt{\frac{\hbar D}{\Delta}} \sim 250 nm \sim L.$$ Here, $D = v_F l / 2$ is the diffusion coefficient. This places our SGS devices at the crossover between long and short diffusive Josephson junctions.

In diffusive junctions (weak links), the values of $I_c$ and of the $I_c R_n$ product are reduced compared to the ballistic case because of scattering. This effect is captured by the Likharev model[23, 24] which we adopt here for the data analysis. The model treats the junction in the diffusive limit as a 1D weak link with vanishing gap in the channel material. With these simplifications and with the mean free path, obtained from the measured $V_g$ dependence of $R_n$, as an input parameter[25], we numerically solved Usadel's equations[26] (with no fitting parameters) to obtain an expression for the temperature dependence of $V_c^{cal} \equiv I_c^{cal} R_n$, (the superscript *cal* refers to calculated values) as a function of doping level. Here $I_c^{cal} = V_c^{cal} / R_n$ is the value of the critical current obtained from the numerical solution. Figure 2a shows the comparison with the experimental results near the DP and at $V_g = 40$ V. The overall temperature dependence of the calculated and measured values of $I_c$ are in qualitative agreement but their magnitude is consistently larger than the measured values (~ 1.5 and 2.5 times larger near the DP and at $V_g = 40$ V respectively). This discrepancy can be attributed to "premature" switching induced by fluctuation from thermal and electro-magnetic noise[27, 28]. The mean reduction in critical current due to premature switching can be estimated in the limit $E_J \gg E_{fl}$, as $\langle \Delta I_c \rangle \sim I_c \left[ \frac{E_{fl}}{2E_J} ln\left(\frac{\omega_p \Delta t}{2\pi}\right) \right]^{2/3}$. Here $E_J = \frac{\hbar I_c}{2e}$ is the Josephson energy, $E_{fl}$ is a characteristic fluctuation energy, $\Delta t \sim 10^2$-$10^3$ s the measurement time, $\omega_p = \sqrt{\dfrac{2eI_c}{\hbar C}} \sim 10^{11}\,\text{s}^{-1}$ is the plasma frequency of the junction and $C \sim 2\times 10^{-13}$ F is the effective capacitance estimated from the RCSJ model[18]. Assuming additive thermal and radiation noise $E_{fl} \sim k_B(T+T_{EM})$, with $T_{EM}$ an effective temperature increase due to the radiation energy, we found that a noise temperature, $T_{EM} \sim 300$ mK, gives good agreement with our data at low temperatures for all doping levels and for all samples measured as illustrated in Figure 2b. $V_c \equiv I_c(V_g) R_n(V_g)$ is plotted as a function of gate voltage. This is compared to $V_c^* = I_c^*(V_g) R_n(V_g)$ (dashed line) where $I_c^* = I_c^{cal} - \langle \Delta I_c \rangle$. We note that $V_c^{cal}$ is almost independent of $V_g$ as expected because the mean free path depends only weakly on $V_g$. The observed "V" shaped dependence is mostly due to fluctuation effects, which are more significant near the DP where $I_c$ is smallest. As shown in figure 2b, this model indeed yields quantitative agreement with the data: the calculated results fit the data throughout the entire range of $V_g$ by assuming a single value for $T_{EM}$ which is consistent with our experimental setup and level of shielding. By contrast, in order to achieve agreement with the predictions for a ballistic junction, one would have to assume noise temperatures that are not only unrealistically high but that would have to depend on gate voltage: 4 K at the Dirac point and 50 K at $V_g = 30$ V. Furthermore, though the ballistic model also predicts a "V" shaped $V_g$ dependence of $I_c R_n$ at low doping levels, the feature ($\dfrac{\mu L}{\hbar \upsilon} \sim 5$, where $\mu$ is the chemical potential and $\upsilon$ is the Fermi velocity) is 2 orders of magnitude narrower than what is observed.

Remarkably, the diffusive model works not only for our sample, but also for published data from other groups as shown in figure 2b. The results of ref. 5, are fit with the same model by assuming a constant noise temperature of ~ 30 mK, consistent with the more stringent shielding conditions in those experiments.

While the critical current is a powerful tool to investigate the interplay between normal and superconducting electrons, a direct comparison with theory is not straightforward. As illustrated above this is due to fluctuations in the superconducting phase that depend on the electromagnetic environment in which the junction is embedded. This can greatly reduce the value of the observed supercurrent as compared to theoretical predictions for idealized situations. By contrast, the *IVC* in the resistive regime are not affected by this noise.

The most prominent feature of the *IVC* for SNS junctions is the appearance of the so-called subharmonic gap structure, which consists of a series of conductance maxima at voltages *2Δ/ne*, where n is an integer and *Δ* is the energy gap of the electrodes as a result Multiple Andreev Reflections. In the case of the relativistic electrons in undoped graphene the Andreev reflections are expected to be specular (SAR)[14]. This leads, in SGS ballistic junctions, to a strong gate dependence of the normalized conductance maxima of the MARs[15]. This is in contrast to the constant value seen in standard SNS junctions. Fig. 3a illustrates the development of pronounced minima in the bias-voltage dependence of the differential resistance due to MAR. The first 4 MAR oscillations are indicated by the dotted lines. For all our samples, the first 4~6 oscillations are easily identified, indicating high transparency of the SN interfaces. The temperature dependence of the sub-gap features provides a measure of *Δ(T)*. In Fig. 3b we plot the bias dependence of the normalized conductance for three gate voltages (this data was taken on another sample S022207 with $L = 220$ nm, $W = 2.8$ μm and $R_{n\_max} = 465$ Ω). Comparing these MAR features with predictions by the diffusive junction model, we find good agreement with the theory[29] for junctions with L/ξ ratio between 1 and 2. This yields a superconducting coherence length of 150 ~ 300 nm, corresponding to a mean free path of 10 ~ 30 nm, in good agreement with our measured values. The agreement holds for all our samples as well as for published data taken on samples from ref.5 . This is illustrated in the inset of Figure 3b where the measured normalized differential conductance at the first subgap peak is plotted against the ratio L/ ξ and compared to the theoretical values[29]. All reported data points fall nicely onto the theoretical curve.

Contrary to the expected gate dependence for SAR in ballistic junctions, the normalized MARs features show no gate dependence within experimental error. The slight reduction in the value of the maximum conductivity of the first MAR peak at the DP is readily explained in terms of the reduced mean free path there[29]. It is likely that the absence of the SAR in the SGS junctions is not intrinsic but rather a consequence of the poor screening afforded by the low carrier density and the low dimensionality of the graphene. Thus any charge inhomogeneity in the substrate or above the graphene link can lead to the formation of electron-hole puddles which inevitably broaden the DP. Indeed, as was shown by SET scanning microscopy[19] these puddles are quite pronounced and lead to a distribution of Fermi levels near the DP corresponding to ~ 2 V in gate modulation, or in terms of the Fermi level variation $\delta E_f = \hbar v_F \sqrt{\dfrac{\varepsilon \varepsilon_0 \delta V_g \pi}{ed}} \sim 45 meV \sim$ ($>>\Delta=0.12meV$). Similar results were obtained in magneto-transport[20] where strong enhancement of longitudinal resistivity and suppression of Hall resistivity were observed near the DP. If the SAR are to be observed, the spread of the DP cannot exceed the energy scale of the superconducting gap. In other words the required charge uniformity corresponds to a gate control of $V_g$ < 0.1 mV (for Al leads) and $V_g$ < 0.1 V (for HTC superconductors). These conditions are not compatible with present fabrication techniques of SGS junctions.

The experiments described here demonstrate that the transport in SGS junctions fabricated on SiO$_2$ substrates is diffusive, with mean free paths (10 ~ 50 nm) much shorter than the lead separation. This limitation is probably not intrinsic since the mean free paths attained in some graphene devices with non superconducting leads and with large (micron size) lead separation can be up to an order of magnitude longer[21]. It is possible that the invasive presence of the Al/Ti leads and the poor screening of the 2D carriers are crucial limiting factors. Thus, improving the design and using different lead and substrate materials may in the future give access to the SAR. However, even within present sample fabrication techniques, the gate tunability of the SGS and its almost ideal interfaces make it a promising candidate for superconducting circuit applications. At the same time these junctions are a powerful tool for exploring the phase diagram of the Josephson Effect with one single device covering the various regimes from underdamped to overdamped phase diffusion[10].

We thank J. Wei, J. Sanchez, G. Li, Z. Chen for enlightening discussions, S.W. Cheong for use of AFM, M. Gershenson for use of E-beam lithography system, V. Kiryukhin and A. F. Hebard for providing the HOPG crystals. Work supported by DOE DE-FG02-99ER45742; NSF-DMR-0456473 and ICAM.

**Figure Captions.**

**Figure 1.** Transport characteristics of SGS junction.
(a) Main panel: current voltage characteristics showing the Josephson current state at T = 200 mK. Upper inset - optical image of a typical device. Lower inset - numerically differentiated IV curves as a function of magnetic field. The yellow line separating the Josephson state at low currents (blue region) from the normal state at high currents (red region) exhibits the oscillating field dependence of the switching current typical of the Frauenhofer pattern.
(b) Gate dependence of the voltage across a junction in the Josephson state carrying a constant current of 800 nA. The switching to the normal state at low gate voltages occurs when the applied current exceeds the critical current of the junction.
(c) Gate voltage dependence of the resistance in the normal state at T=200mK. A small magnetic field was applied to suppress the superconductivity of the leads. Inset – mean free path calculated from the transport data. The solid (red) dot and the open (blue) dot indicate estimates of the mean free path near the DP and at $V_g = 35$ V, respectively.
(d) Numerically differentiated IV curves as a function of gate voltage. The center blue area (color online) corresponding to the Josephson state is separated from the normal state, purple area, by the switching current represented by the bright (yellow) line.

**Figure 2.** Doping and temperature dependence of the Josephson effect.
(a) Comparison of the measured temperature dependence of the switching current, $I_c$, with calculated values, $I_c^{cal}$, (without fluctuations).
(b) Comparison of measured gate dependence of $V_c = I_c R_n$ (red curve) with calculated values from Likharev's model and with corrections for premature switching $V_c^* = I_c^* R_n$ (black curve). Inset - ratio of experimental and theoretical values $V_c / V_c^*$, for two of our samples: S032007, S02207, and for data obtained from ref.5 .The black line is a guide to the eye.

**Figure 3.** Multiple Andreev Reflections.
(a) Temperature dependence of the MAR. The dotted lines are guides to the eye for the first four sub-gap oscillations ($2\Delta/n$, n=1,2,3,4). Curves of resistance versus. bias voltage taken for a sequence of equally spaced temperatures (between 200 mK and $T_c$) are shifted along the y-axis for clarity. The extra peaks at high bias (seen in the curve just below $T_c$ ) signal the superconducting to normal transition of the leads (when the applied current equals the lead critical current) because this is not a true four lead measurement.

(b) Comparison of the normalized sub-gap features at different doping levels for sample S022207. Main panel: Bias dependence of normalized differential conductance.

Inset: normalized differential conductance as a function of $L/\xi$. The black squares represent theoretical values for diffusive junctions from ref.29. The triangles correspond to measured values for samples S032007, S02207, and for the data obtained from ref.5. A similar estimate for the single layer graphene data in ref.6 was not possible because of the absence of clear Andreev reflections. The dotted line is a guide to the eye.

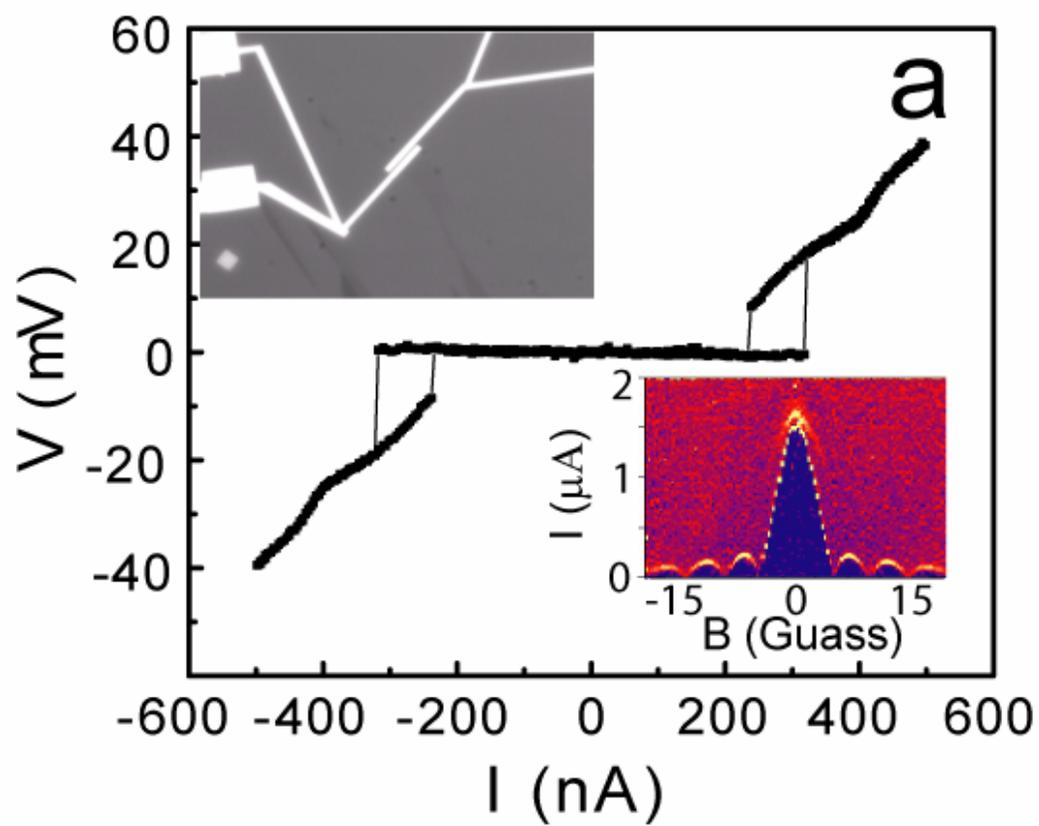

Figure 1(a)

Size x 3

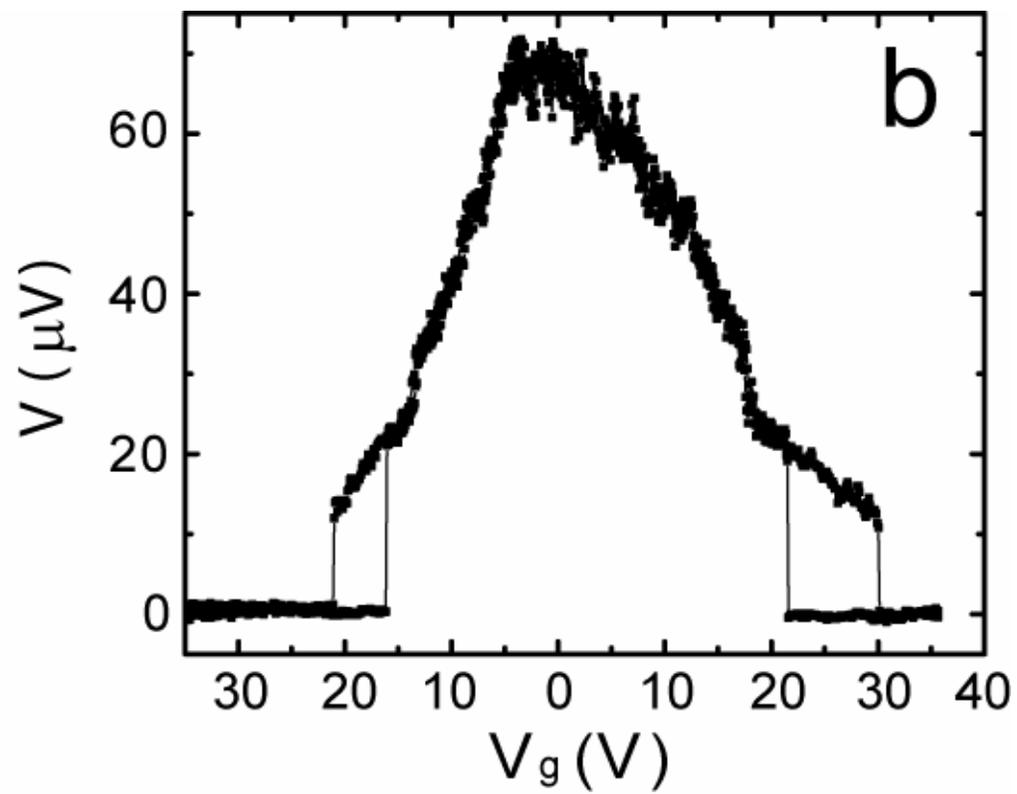

Figure 1(b)
Size x 3

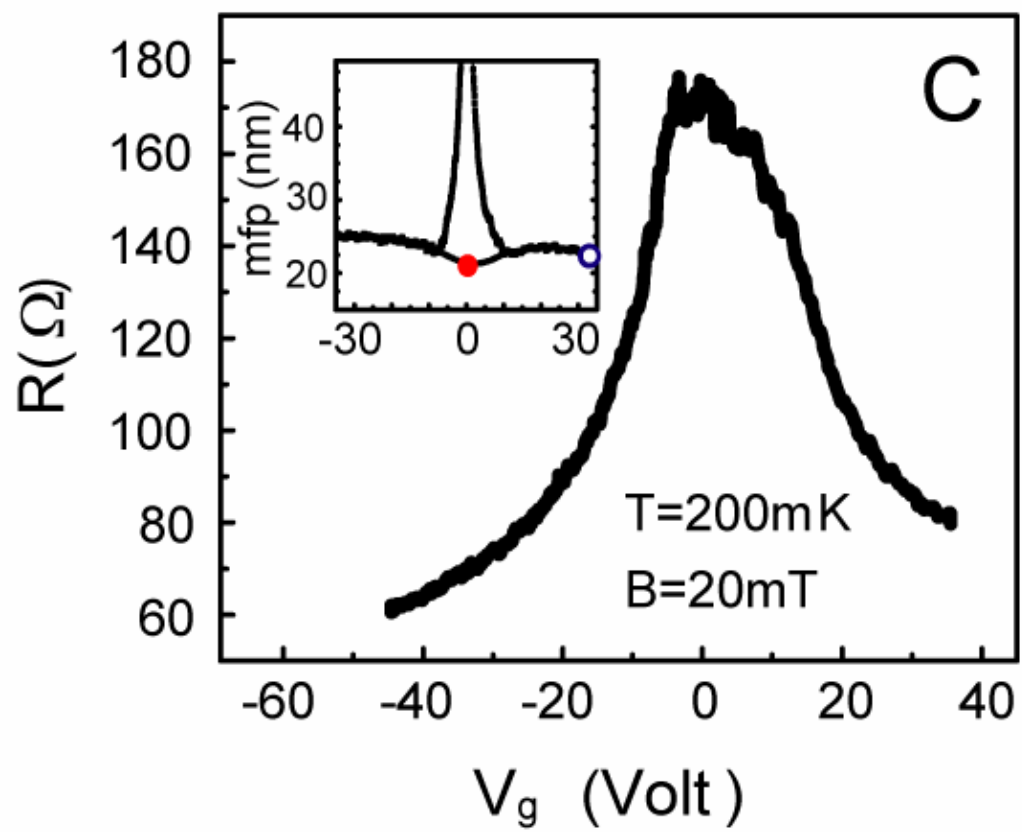

Figure 1(c)
Size x 3

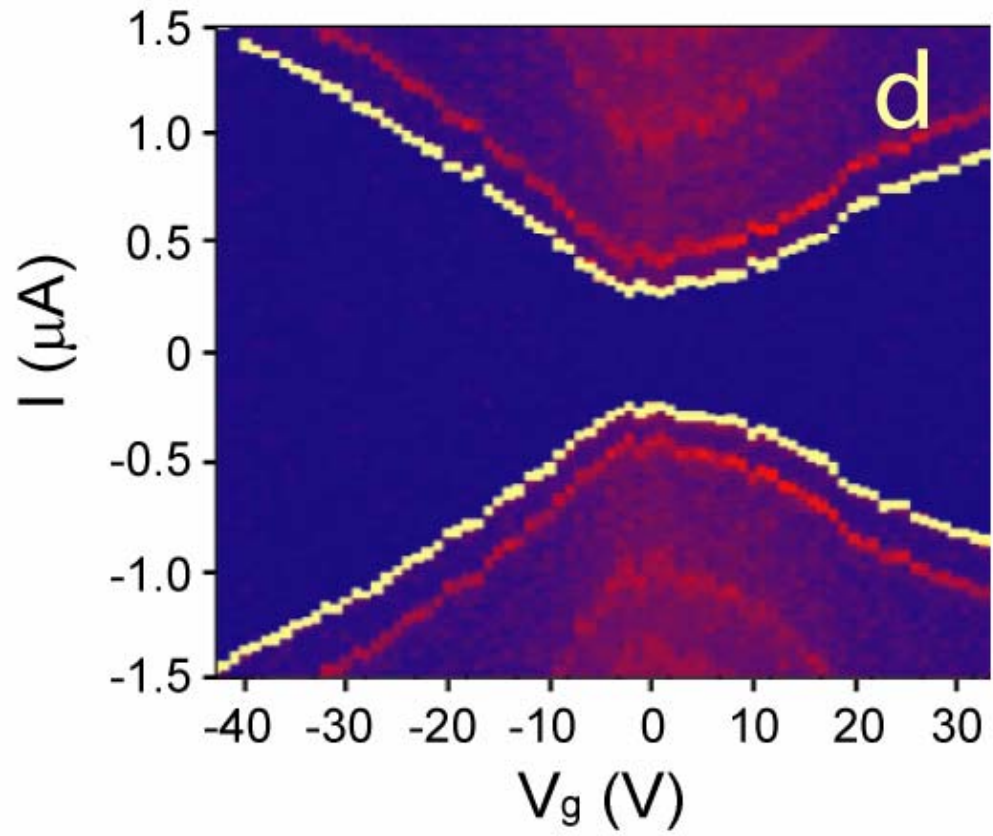

Figure 1(d)

Size x 3

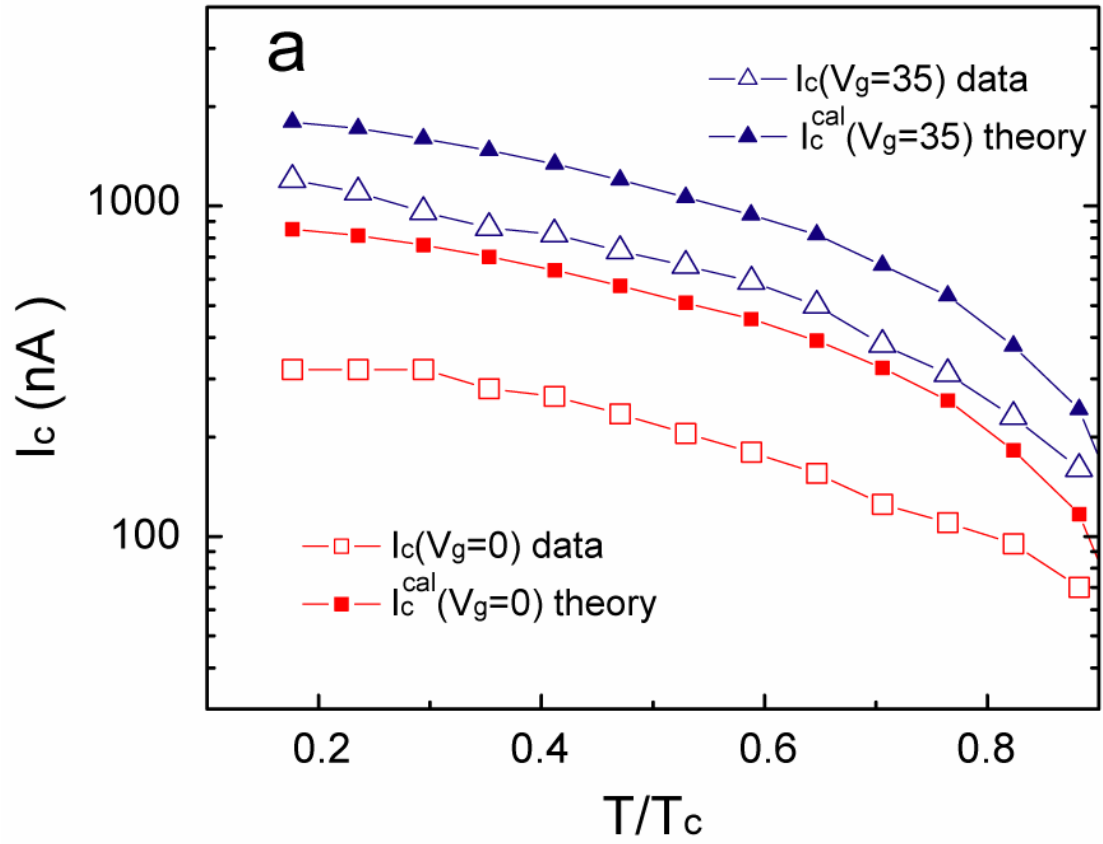

Figure 2(a)
Size x 2

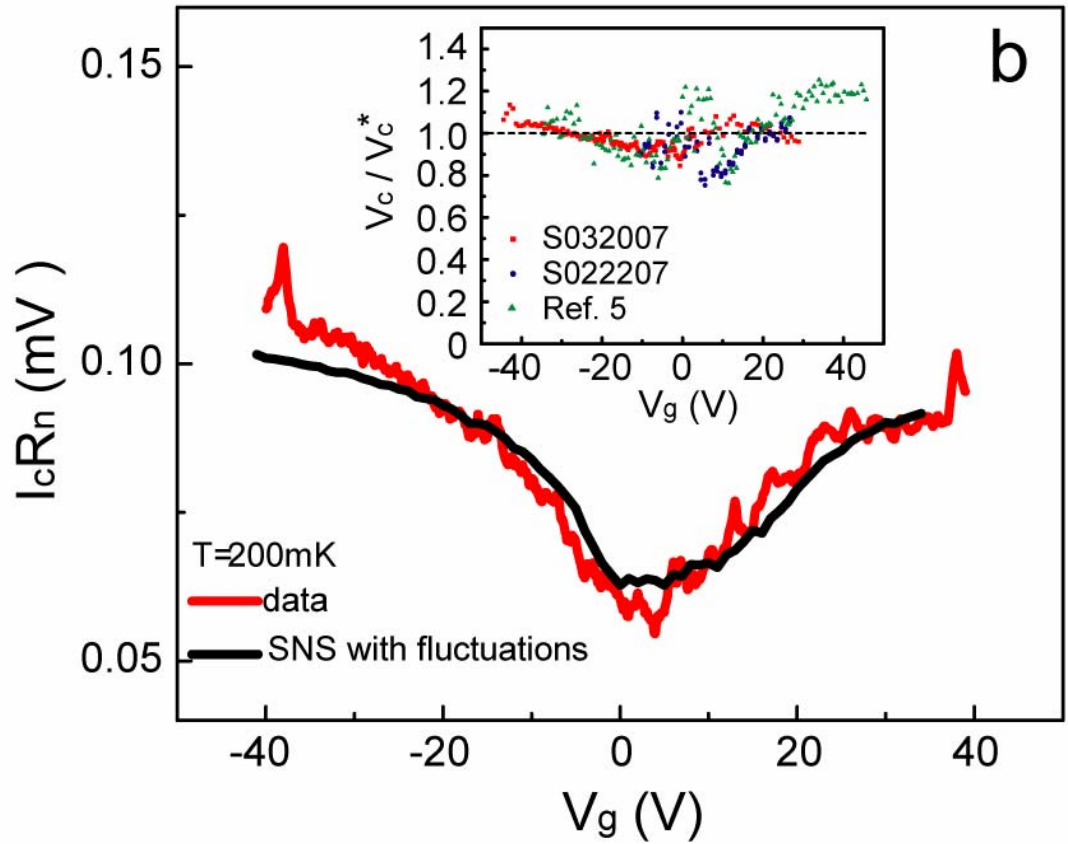

Figure 2(b)

Size x 2

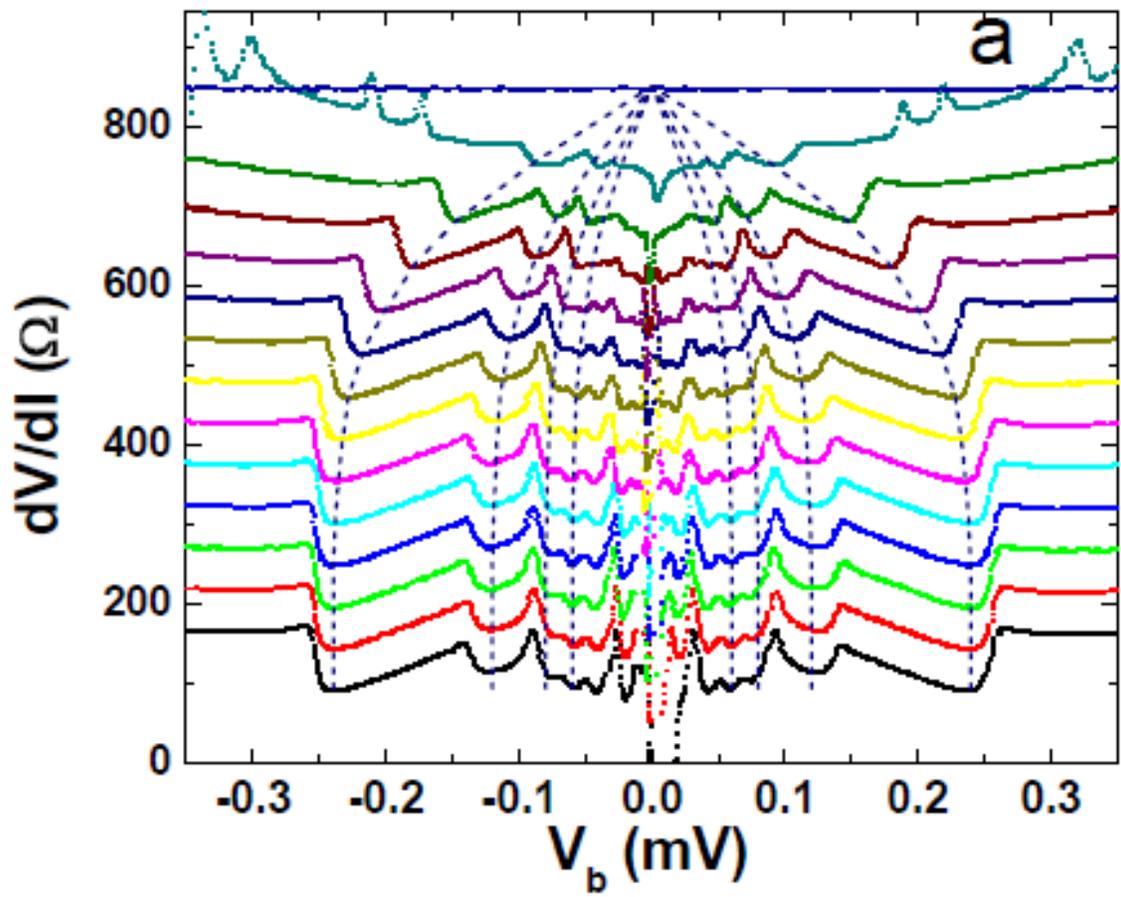

Figure 3(a)

Size x 2

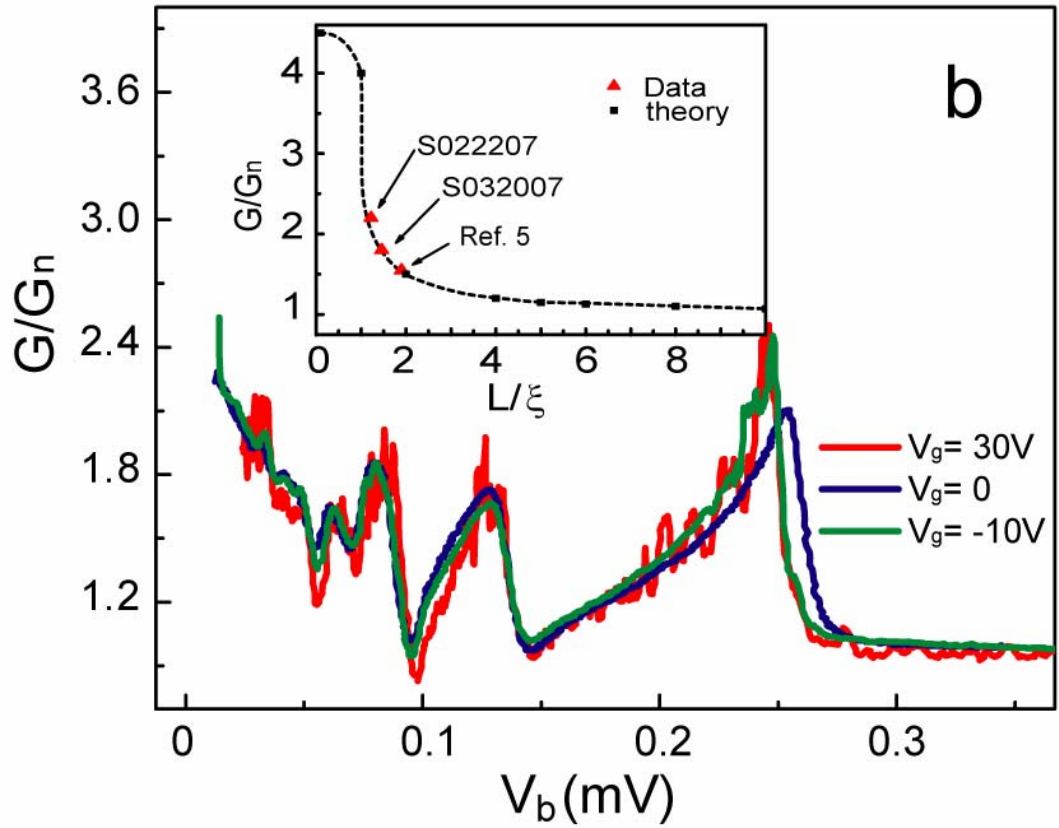

Figure 3(b)

Size x 2